\documentclass[aip, reprint, amsmath,amssymb,citeautoscript]{revtex4-1}
\usepackage{graphicx}
\usepackage{dsfont}
\draft % marks overfull lines with a black rule on the right

\begin{document}

%\preprint{}

\title{Good and bad predictions: Assessing and improving the replication of chaotic attractors by means of reservoir computing} %Title of paper



%\author{}
%\thanks{}
%\affiliation{}

\author{Alexander Haluszczynski}
\email{alexander.haluszczynski@gmail.com}
\affiliation{Ludwig-Maximilians-Universit\"at, Department of Physics, Schellingstra{\ss}e 4, 80799 Munich, Germany}
\affiliation{risklab GmbH, Seidlstra{\ss}e 24, 80335, Munich, Germany}
\author{Christoph R{\"a}th}
\email{christoph.raeth@dlr.de}
\affiliation{Deutsches Zentrum f{\"u}r Luft- und Raumfahrt, Institut f{\"u}r Materialphysik im Weltraum,
M{\"u}nchner Str. 20, 82234 Wessling, Germany}


\date{\today}

\begin{abstract}

The prediction of complex nonlinear dynamical systems with the help of machine learning techniques has become increasingly popular. In particular, reservoir computing turned out to be a very promising approach especially for the reproduction of the long-term properties of a nonlinear system. Yet, a thorough statistical analysis of the forecast results is missing. Using the Lorenz and R{\"o}ssler system we statistically analyze the quality of prediction for different parametrizations - both the exact short-term prediction as well as the reproduction of the long-term properties (the ``climate") of the system as estimated by the correlation dimension and largest Lyapunov exponent. We find that both short and longterm predictions vary significantly among the realizations. Thus special care must be taken in selecting the good predictions as predictions which deliver better short-term prediction also tend to better resemble the long-term climate of the system. Instead of only using purely random Erd{\"o}s-Renyi networks we also investigate the benefit of alternative network topologies such as small world or scale-free networks and show which effect they have on the prediction quality. Our results suggest that the overall performance with respect to the reproduction of the climate of both the Lorenz and R{\"o}ssler system is worst for scale-free networks. For the Lorenz system there seems to be a slight benefit of using small world networks while for the R{\"o}ssler system small world and Erd{\"o}s-Renyi networks performed equivalently well. In general the observation is that reservoir computing works for all network topologies investigated here.

\end{abstract}

%\pacs{XXX}% insert suggested PACS numbers in braces on next line

\maketitle %\maketitle must follow title, authors, abstract and \pacs

\begin{quotation}

The application of machine learning techniques to various fields in science and technology yields very promising and fast advancing results. However, the robustness of these methods is a critical aspect that is often not adequately addressed. Particularly when trying to predict complex nonlinear systems -- here by using a recurrent neural network based approach called reservoir computing -- it is very useful to know how likely it is to end up with a \textit{good} prediction and how different the results can be in terms of quality. In our context a good prediction is achieved when the predicted trajectory matches those of the actual system in the short-term while reproducing its statistical properties in the long-term. In order to thoroughly investigate the prediction quality we run our analysis not only using a single prediction but on many realizations which are based on the same parameters but different random number seeds. As a result we find strong variability among the quality of the predictions, indicating that robustness seems to be an issue and show the effect of varying the network topology of the reservoir. 

\end{quotation} 

\section{Introduction}
In the recent years the use of machine learning (ML) techniques has not only become increasingly important in research but also popular in media, public perception and businesses. The euphoric application in all possible areas, however, carries the risk of misinterpreting the results if deeper methodological knowledge is lacking. This is reminiscent of the situation in the late 1980s and early 1990s when chaos was a hot topic in the scientific community. In the absence of adequate statistics analysis, many systems have been erroneously categorized as being chaotic on the basis of e.g. assessing the attractor dimensions by single measurements of short and noisy time series. 
Only after Theiler et al. \cite{theiler92} introduced the concept of surrogate data the errors of the nonlinear measures for a given data set could be assessed, and it turned out that many claims of chaos had to be rejected. The lesson learned is that in the absence of a proper (linear) model of the underlying process 
credible results can only be obtained by applying thorough statistical analyses involving averaging over a large number of realizations of simulations or surrogates.\\
In recent years the use of reservoir computing for quantifying and predicting the spatiotemporal dynamics of nonlinear systems has attracted much attention \cite{lu17, pathak2017using,pathak18a,pathak18b, zimmermann18,carroll18,lu2018attractor,antonik18}.
Many of the achievements  --  be it e.g. the cross-prediction of variables in two-dimensional excitable media \cite{zimmermann18} or  the reproduction of the 
spectrum of Lyapunov exponents in lower dimensional (Lorenz or R{\"o}ssler) and higher dimensional (Kuramoto-Sivashinsky) systems \cite{lu17, pathak2017using,pathak18a} -- are impressive and guide the way to a range of new applications of ML in complex systems research. However,  all results shown until now are based on a single or few realizations of reservoir computing.
It is thus  so far impossible to judge the robustness of the results on e.g. variations of the set of random variables specifying the reservoir. Here, we perform the first thorough statistical analysis of predicting short and long-term behaviour of nonlinear time series by means of reservoir computing.\\
The heart of reservoir computing is -- as the name already says -- a so-called reservoir, which consists of $D_r$ nodes that are sparsely connected with each other. The nodes are supposed to yield a proper ``echo state'' to a given input, which is then transferred to the output layer. That's why most types of reservoir networks are often called ``echo state networks (ESN)''. Beginning with the first introduction of ESNs by Maass and Jaeger \cite{maass02, jaeger04} the reservoir has typically been modelled as a random 
Erd{\"o}s R\'enyi network, where two nodes are connected with as certain probability $p$. However, the groundbreaking works of Watts and Strogatz \cite{watts1998collective},  Albert and Barabasi \cite{albert02}. and many others have shown that random networks are far from being common in physics, biology, finance or sociology.  Rather,   
more complex networks like scale-free, small world or intermediate forms of networks \cite{broido18,gerlach19} with intriguing new properties are most often found in real world applications. Having this in mind it seems natural to ask whether also for reservoir computing the topology of the network has a significant influence on the prediction results \cite{cui2012architecture}. As a first step we use the three aforementioned prototypical classes of networks as reservoir and compare them regarding their ability of short and long-term prediction of time series.\\
The paper is organized as follows: Section~\ref{Methods} introduces reservoir computing and the methods used in our study. In section~\ref{Results} we present the main results obtained from the statistical analysis of the prediction results as well as studying different reservoir topologies. Our summary and the conclusions are given in section~\ref{Conclusion}.

%%%%%%%%%%%%%%%%%%%%%%%%%%%%%%%
\section{Methods}
\label{Methods}
%%%%%%%%%%%%% A Lorenz and Roessler System
\subsection{Lorenz and R{\"o}ssler system}
As in Pathak et al. \cite{pathak2017using} and Lu et al. \cite{lu2018attractor} we use the Lorenz system \cite{lorenz1963deterministic} as an example for replicating chaotic attractors using reservoir computing. It has been developed as a simplified model for atmospheric convection and exhibits chaos for certain parameter ranges. The standard Lorenz system, however, is symmetric in $x$ and $y$ with respect to the transformation $x \rightarrow -x$ and $y \rightarrow -y$. This can be an issue for example when trying to infer the $x$ and $y$ dimension from knowledge of the $z$ dimension as outlined in Lu et al. \cite{lu2017reservoir}. In order to study a more general example we would like to modify the Lorenz system such that this symmetry is broken. This can be achieved by adding the term $x$ to the $z$-component which then reads:
\begin{eqnarray}
\begin{aligned}
\dot x &= \sigma (y-x) \\
\dot y &= x (\rho-z)-y \\
\dot z &= x y - \beta z + x \ .
\label{eq:lorenz}
\end{aligned}
\end{eqnarray} 
We use the standard parameters $\sigma = 10 , \beta = 8/3$ and $\rho = 28$. This system is referred to as modified Lorenz system. The equations are solved using the 4th order Runge-Kutta method with a time resolution $\Delta t = 0.02$. 

In addition to the Lorenz system we ran the same analysis also on the R{\"o}ssler system \cite{rossler1976equation} which equations read
\begin{eqnarray}
\begin{aligned}
\dot x &= -y-z \\
\dot y &= x+ay \\
\dot z &= b+z(x-c) \ , 
\label{eq:roessler}
\end{aligned}
\end{eqnarray} 
where we use the parameters $a=0.5$ , $b=2.0$ and $c=4.0$. Again this is a $D=3$ dimensional chaotic system but said to be less chaotic than the Lorenz attractor. Thus it is an interesting object to study in particular when it comes to the short-term prediction capabilities of reservoir computing. For the R{\"o}ssler system the time resolution is chosen to be $\Delta t = 0.05$ in order to ensure a sufficient manifestation of the attractor in the $t_{train} = 5000$ training time steps. 

%%%%%%% Reservoir computing
\subsection{Reservoir Computing} 
Reservoir computing is a machine learning technique that falls into the category of artificial recurrent neural networks. The core of the model is a network called $reservoir$ which --- in contrast to feedforward neural networks --- exhibits loops. This means that past values feed back into the system and thus allow for dynamics \cite{lukovsevivcius2009reservoir, genccay1997nonlinear}. In order to complete the task of predicting time series, the ability to capture dynamics is essential. Moreover, reservoir computing has a powerful advantage: While in other methods the network itself is dynamical, here the training is based only on the linear output layer and therefore allows for large reservoir dimensionality while still being computationally feasible \cite{lu2018attractor}.

In our implementation we stick to the setup used by Pathak et al. \cite{pathak2017using} which works as follows. We have an input signal $\textbf{u}(t)$ with dimension $D$ that we would like to feed into a reservoir $\textbf{A}$. The reservoir is chosen to be a sparse Erd{\"o}s-Renyi random network with $D_{r}=300$ nodes and $p=0.02$ \cite{erdos1959random}. Here $p$ describes the probability of connecting two nodes which then leads to an unweighted average degree of $d=6$. To obtain the weighted network we then replace all nonzero elements of the adjacency matrix by independently and uniformly drawn numbers from $[-1,1]$. It is important to highlight that the network itself is static and thus does not change over time. In order to feed the lower dimensional input signal $\textbf{u}(t)$ into the reservoir, an $D_{r} \times D$ input function $\textbf{W}_{in}$ is required. The entries of $\textbf{W}_{in}$ are chosen here to be uniformly distributed random numbers within the range of the nonzero elements of the reservoir. 

A key property of the system are its $D_{r} \times 1$ reservoir states $\textbf{r}(t)$ which represent the scalar states of the nodes of the reservoir network. We initially assume $r_{i}(t_{0})=0$ for all nodes and update the reservoir states in each time step according to the equation
\begin{eqnarray}
\scalebox{0.90}[1]{$\textbf{r}(t+ \Delta{t}) = \alpha \textbf{r}(t)  + (1-\alpha) tanh(\textbf{A}\textbf{r}(t) + \textbf{W}_{in} \textbf{u}(t))$} \ .
\label{eq:updating}
\end{eqnarray} 
As in Pathak et al. \cite{pathak2017using} we set $\alpha = 0$ and therefore do not mix the input function with past reservoir states. Now we have a fully dynamical system where the network edges are constant and the states of the nodes are time dependent.

The next step is to map the reservoir states $\textbf{r}(t)$ back to the $D$ dimensional output \textbf{v} given by
\begin{eqnarray}
\textbf{v}(t+ \Delta{t}) = \textbf{W}_{out}(\textbf{r}(t+ \Delta{t}),  \textbf{P}) \ .
\label{eq:output}
\end{eqnarray} 
Here we assume that $\textbf{W}_{out}$ depends linearly on $\textbf{P}$ and reads $\textbf{W}_{out}(\textbf{r},\textbf{P}) = \textbf{P}\textbf{r}$. This means that the output depends only on the reservoir states $\textbf{r}(t)$ and the output matrix $\textbf{P}$ which contains a large number of adjustable parameters - all its elements. Therefore, after acquiring a sufficient number of reservoir states $\textbf{r}(t)$ we have to choose $\textbf{P}$ such that the output $\textbf{v}$ of the reservoir is as close as possible to the known real output $\textbf{v}_{R}$. This process is called training. In general, the task is to find an output matrix \textbf{P} using Ridge regression, which minimizes 
\begin{eqnarray}
\sum^{}_{-T \leq t \leq 0} {\parallel  \textbf{W}_{out}(\textbf{r}(t), \textbf{P}) - \textbf{v}_{R}(t) \parallel}^2 - \beta {\parallel \textbf{P} \parallel}^2 \ ,
\label{eq:minimizing}
\end{eqnarray} 
% Define r (without t)
where $\beta$ denotes the regularization constant, that prevents from overfitting by penalizing large values of the fitting parameter. In this study we choose $\beta=0.01$. The notation, $\parallel \textbf{P} \parallel$ describes the sum of the square elements of the matrix $\textbf{P}$. For solving this problem, we are applying the matrix form of the Ridge regression \cite{hoerl1970ridge} which leads to
\begin{eqnarray}
\textbf{P} = {(\textbf{r}^{T} \textbf{r} + \beta \mathds{1})}^{-1} \textbf{r}^{T} \textbf{v}_{R} \ .
\label{eq:ridgematrix}
\end{eqnarray} 
%explain that r without (t) mean the whole matrix and not the vector r(t)
The notion $\textbf{r}$ and $\textbf{v}_{R}$ without the time indexing denotes matrices where the columns are the vectors $\textbf{r}(t)$ and $\textbf{v}_{R}(t)$ respectively in each time step. In our implementation we chose $t_{train} = 5000$ training time steps while allowing for a washout or initialisation phase of $t_{init} = 100$. During this time we do not "record" the reservoir states $\textbf{r}(t)$, which means that only $4900$ time steps are used for the regression. In order to ensure that $100$ time steps washout is sufficient we also ran the analysis with $1000$ time steps washout and found both results to be equivalent.

After \textbf{P} is determined we can now switch to the prediction mode by giving the predicted state $\textbf{v}(t)$ as input instead of the actual data $\textbf{u}(t)$. The update equation for the network states $\textbf{r}(t)$ then reads
\begin{eqnarray}
\begin{aligned}
\textbf{r}(t+ \Delta{t}) &= tanh(\textbf{A}\textbf{r}(t) + \textbf{W}_{in} \textbf{W}_{out}(\textbf{r}(t),\textbf{P})) \\
                                  &= tanh(\textbf{A}\textbf{r}(t) + \textbf{W}_{in} \textbf{P}\textbf{r}(t))  \ .
\label{eq:updatingprediction}
\end{aligned}
\end{eqnarray} 
This allows us to produce a predicted trajectory of any length by just applying Eq. \ref{eq:output}.

%%%%% C Alternative Network Topologies
\subsection{Alternative network topologies} 
So far it has been standard practice to use purely random Erd{\"o}s-Renyi networks for the reservoir $\textbf{A}$. However, there is a variety of conceivable network topologies that may have an influence on the results. In this study we investigate the use of \textit{Small World} \cite{watts1998collective} and \textit{Scale Free} \cite{barabasi2003scale} networks as an alternative. 

Small World networks are graphs where the distance -- in terms of steps via other nodes -- between any pair of nodes is small. At the same time the clustering coefficient is relatively high which means that neighbouring nodes tend to be connected. This so-called \textit{small world property} is observed in many real world networks such as social networks, electric power grids, chemical reaction networks and neuronal networks \cite{Amaral2000}.In order to have the same average degree $d=6$ as the random Erd{\"o}s-Renyi networks we construct the Small World networks in the following way: First we connect each node with its six nearest neighbours implying periodic boundary conditions. This is equivalent to arranging all nodes as a ring. Then we loop over each edge $x-y$ and rewire it to $x-z$ with probability $p=0.2$, where node $z$ is randomly chosen. 

Scale Free networks are graphs where the distribution of the number of edges per node decays with a power law tail. This is again a property which is observed in many real world networks. For example, the above mentioned electric power grid networks and neuronal networks exhibit not only the small world property but are also scale free. Other examples include the world wide web or networks of citations of scientific papers \cite{Amaral2000}. Again, the network is constructed such that its average degree is $d=6$. For this we use the scale free graph generator of the \textit{NetworkX} package \cite{hagberg2008exploring} with parameters $\alpha=0.285, \beta=0.665, \gamma=0.05, \delta_{in}=0.2, \delta_{out}=0$. Note that in this case the graph is directed while the other two network topologies result in undirected graphs. Here, $\alpha$ sets the probability of adding a new node which is connected to an already existing node which is chosen randomly according to the in-degree distribution while $\gamma$ does the same except that the node is chosen according to the out-degree distribution. In addition, $\beta$ regulates the probability of creating an edge between two existing nodes where one is chosen according to the in-degree distribution and the other node according to the out-degree distribution \cite{bollobas2003directed}. %explain in and out degree distribution

%%%%%%%%%%% Section about measures %%%%%%%%%%%%%%%%%%%%%%%%%%%%%%%%%%%%%%%%%%%%%%
\subsection{Measures of the System} 
% Fig %%%
\begin{figure*}[t!] 
  \begin{center}
    \includegraphics[width=1\linewidth]{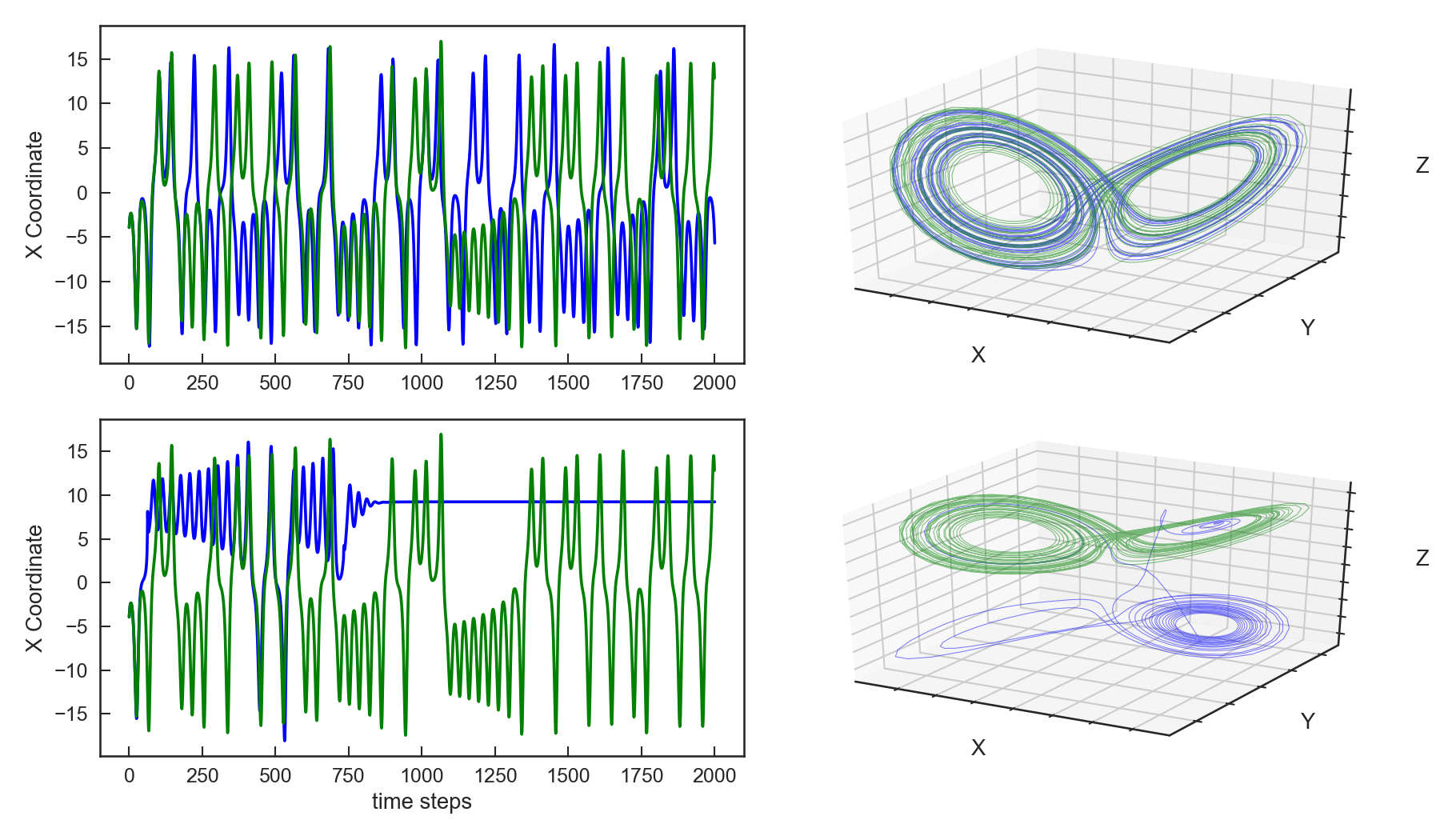}
    \caption{Left: X coordinate of two predicted (blue) trajectories of the Lorenz system plotted over $n=2000$ time steps. The results are compared against the trajectory of the simulated actual Lorenz system (green). The upper plot shows a good realization where both trajectories are overlapping while the lower part shows a bad prediction. Right: Three dimensional attractor for both cases. The spectral radius is $\rho = 0.3$ and random Erd{\"o}s-Renyi networks are used for the reservoir $\textbf{A}$. The correlation dimension for is $\nu = 1.992$ for the upper and $\nu = 0.007$ for the lower realization while the largest Lyapunov exponents are $\lambda_{1} = 0.851$ (upper) and $\lambda_{1} = 0.420$ (lower).}  
    \label{fig:goodbad}
  \end{center}
\end{figure*}

In order to assess the quality of the prediction we are mainly using three different measures. The goal is to adequately address both the exact short-term prediction as well as the long-term reproduction of the statistical properties of the system - its so called climate. 

\subsubsection{Forecast Horizon} 
To quantify the quality and duration of the exact prediction of the trajectory we use a fairly simple measure which we call \textit{forecast horizon}. Here we track the number of time steps during which the predicted and the actual trajectory are matching. As soon as one of the three coordinates exceeds certain deviation thresholds we consider the trajectories as not matching anymore. Throughout our study we use
\begin{eqnarray}
| \textbf{v}(t) - \textbf{v}_{R}(t) | > \boldsymbol{\delta}
\label{eq:fchor}
\end{eqnarray} 
where the thresholds are $\boldsymbol{\delta} = (5, 10, 5)^{T}$ for the Lorenz system and $\boldsymbol{\delta} = (2.5, 2.5, 4)^{T}$ for the R{\"o}ssler system. The values are chosen this way due to the different ranges of the state variables in both systems. The aim is that small fluctuations around the actual trajectory as well as minor detours do not exceed the threshold. Empirically we found that distances between the trajectories become much larger than the threshold values as soon as short-term prediction collapses. 
% Fig %%%
\begin{figure*}[t!] 
  \begin{center}
    \includegraphics[width=1\linewidth]{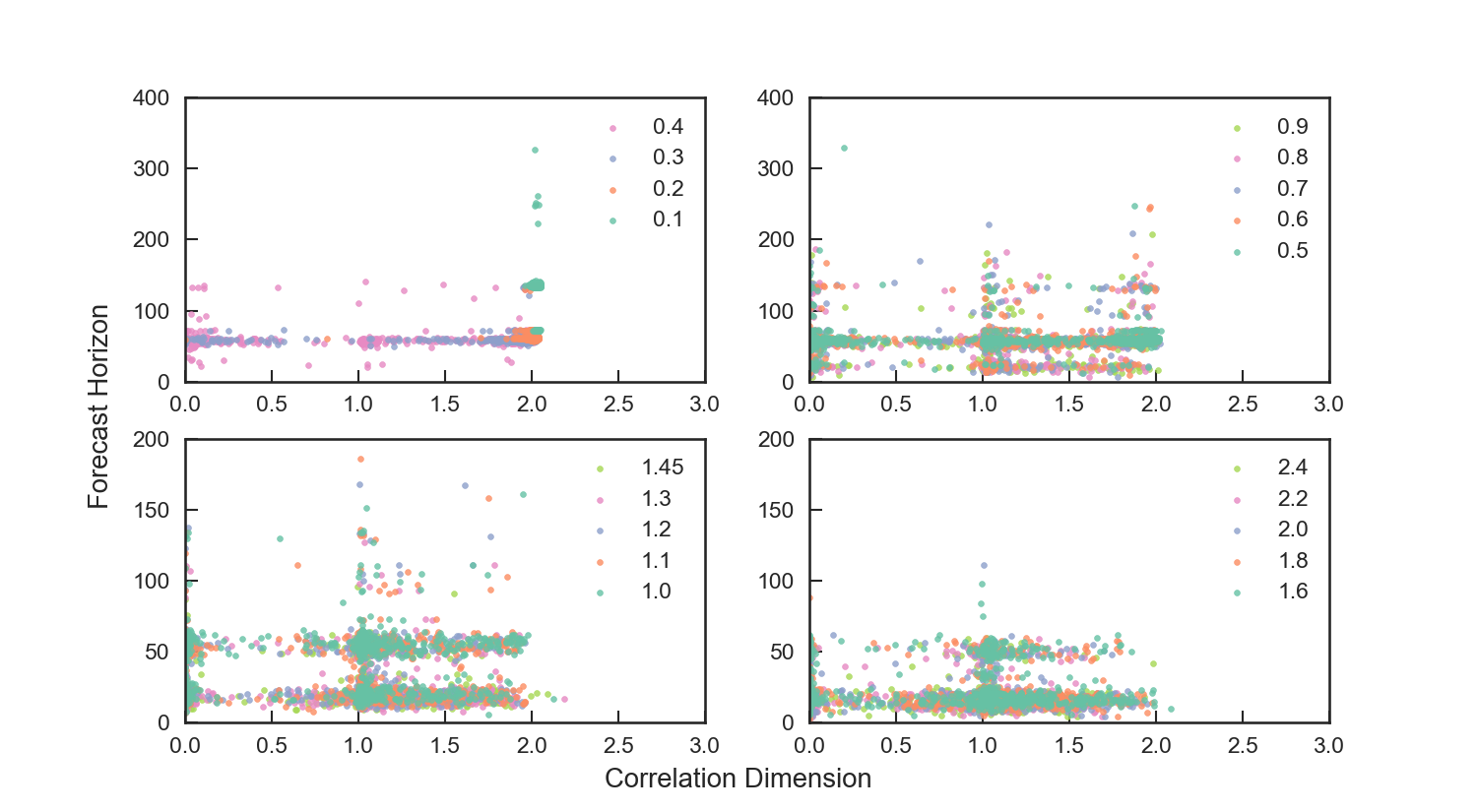}
    \caption{Forecast horizon scattered against the correlation dimension for each of the $N=1000$ predictions of the Lorenz system per spectral radius. Different spectral radii are differentiated by colours. Random Erd{\"o}s-Renyi networks are used for the reservoir $\textbf{A}$.}
    \label{fig:corrfc}
  \end{center}
\end{figure*}

\subsubsection{Correlation Dimension} 
One important characteristic of the long-term properties of the system is its structural complexity. This can be quantified by calculating the correlation dimension which measures the dimensionality of the space populated by the trajectory \cite{grassberger1983measuring}. It is based on the correlation integral 
\begin{eqnarray}
\begin{aligned}
C(r) &= \lim\limits_{N \rightarrow \infty}{\frac{1}{N^2}\sum^{N}_{i,j=1}\theta(r- | \textbf{x}_{i} - \textbf{x}_{j}  |)}      \\
       &= \int_{0}^{r} d^3 r^{\prime} c(\textbf{r}^{\prime}) \ ,
\label{eq:corrintegral}
\end{aligned}
\end{eqnarray} 
which describes the mean probability that two states in phase space are close two each other at different time steps. The condition \textit{close to} is met if the distance between the two states is less than the threshold distance $r$. $\theta$ represents the Heaviside function while $c(\textbf{r}^{\prime})$ denotes the standard correlation function. For self-similar strange attractors the following power-law relationship holds in a range of $r$:
\begin{eqnarray}
C(r) \propto r^{\nu} \ .
\label{eq:corrdim}
\end{eqnarray} 
The \textit{correlation dimension} is then measured by the scaling exponent $\nu$. We use the Grassberger Procaccia algorithm \cite{grassberger83a} to calculate the correlation dimension of our trajectories. This approach is purely data driven and therefore does not require any knowledge about the system. 

%%%%%%
\subsubsection{Largest Lyapunov Exponent} 
Another aspect of the long-term behaviour is the temporal complexity of the system. When dealing with chaotic systems, looking at its Lyapunov exponents is an obvious choice. The Lyapunov exponents $\lambda_{i}$ describe the average rate of divergence of nearby states in phase space and thus measure sensitivity to initial conditions. For each dimension in phase space there is one exponent. If the system exhibits at least one positive Lyapunov exponent it is classified as chaotic while the magnitude of the exponent quantifies the time scale on which the system becomes unpredictable \cite{wolf1985determining, shaw1981strange}. Therefore it is sufficient for our analysis to determine only the largest Lyapunov exponent $\lambda_{1}$
\begin{eqnarray}
d(t) =  C e^{\lambda_{1} t} \ ,
\label{eq:lyapunov}
\end{eqnarray} 
which makes the task computationally easier. Here, $d(t)$ is the average distance or separation of the initially nearby states at time $t$ and $C$ is a constant that normalizes the initial separation. To calculate the largest Lyapunov exponent we use the Rosenstein algorithm \cite{rosenstein1993practical}.

%%%%%%%%%%%%%%%%%%%%%%%%%%%%%%%
\section{Results}
\label{Results}
% Motivation and description what we are doing
Although machine learning techniques and reservoir computing in particular have become increasingly popular, a thorough statistical analysis of the forecast results is yet missing. Therefore we found it insightful to not only perform one single prediction where \textit{prediction} means forecasting the trajectory for $t_{prediction} = 10000$ time steps. As there are random numbers involved in the construction of the reservoir $\textbf{A}$ as well as the input function $\textbf{W}_{in}$ we can run the prediction with $N=1000$ different random number seeds while keeping all other parameters of the network constant. Therefore, for different seeds both $\textbf{A}$ and $\textbf{W}_{in}$ will vary. This allows us to gain a statistical view on the quality of the prediction instead of analysing only single realizations. 
% Fig %%%
\begin{figure*}[t!] 
  \begin{center}
    \includegraphics[width=1\linewidth]{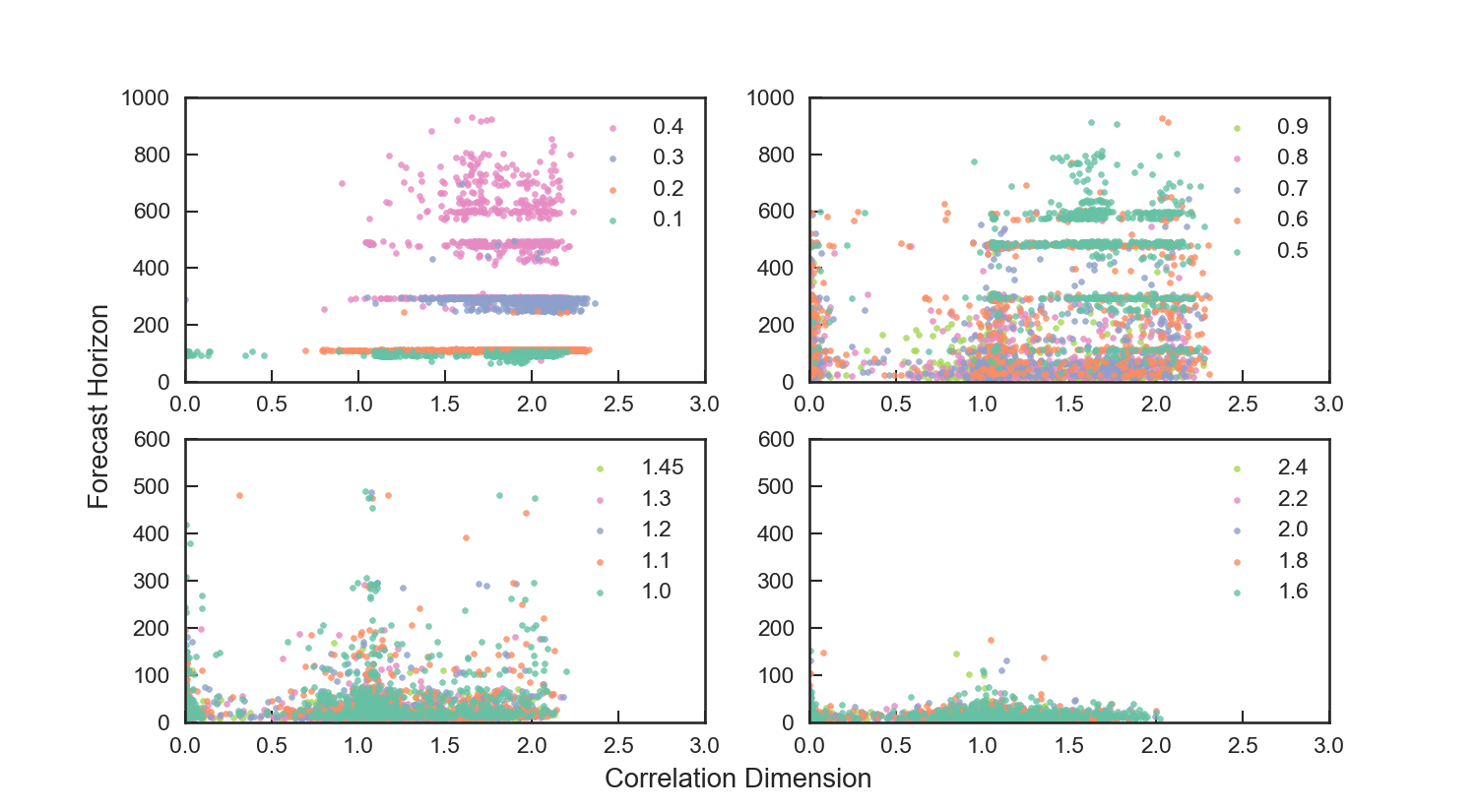}
    \caption{Forecast horizon scattered against the correlation dimension for each of the $N=1000$ predictions of the R{\"o}ssler system per spectral radius. Different spectral radii are differentiated by colours. Random Erd{\"o}s-Renyi networks are used for the reservoir $\textbf{A}$.}
    \label{fig:corrfcRoessler}
  \end{center}
\end{figure*}
In addition to the parameters mentioned in Section \ref{Methods} there is one more parameter that we can tune. This is the spectral radius $\rho$ of the adjacency matrix of the reservoir $\textbf{A}$ which is defined as its largest absolute eigenvalue 
\begin{eqnarray}
\rho(\textbf{A}) =  max \{|\lambda_{1}|, ..., |\lambda_{D_{r}}|\} \ 
\label{eq:spectralradius}
\end{eqnarray} 
and reflects some kind of average degree of the network. We can adjust the spectral radius by 
\begin{eqnarray}
\textbf{A}^{*} =  \frac{\textbf{A}}{\rho(\textbf{A})} \rho^{*} \ , 
\label{eq:spectralradiuschange}
\end{eqnarray} 
where $ \rho^{*}$ is the desired spectral radius. Note that $\lambda_{i}$ here denote the eigenvalues of the adjacency matrix of the reservoir $\textbf{A}$ - not to be confused with the Lyapunov exponent in Eq. \ref{eq:lyapunov} which is commonly called $\lambda$ as well. Other studies showed results for particular values of $\rho$ such as Pathak et al. \cite{pathak2017using} e.g. claiming that the prediction using a spectral radius of $\rho = 1.2$ accurately resembles the long-term climate of the system while the same setup with $\rho = 1.45$ does not. To possibly reproduce these results and to assess the best ranges for the spectral radius we ran the reservoir computing with $N=1000$ different random number seeds for each spectral radius 
$\rho_{i}^{*} \in \{0.1, 0.2, 0.3, 0.4, 0.5, 0.6, 0.7, 0.8, 0.9, 1.0, 1.1, 1.2, 1.3, 1.45 \\
, 1.6, 1.8, 2.0, 2.2, 2.4\}$ while all other parameters remain constant.
\begin{figure*}[t!] 
  \begin{center}
    \includegraphics[width=1\linewidth]{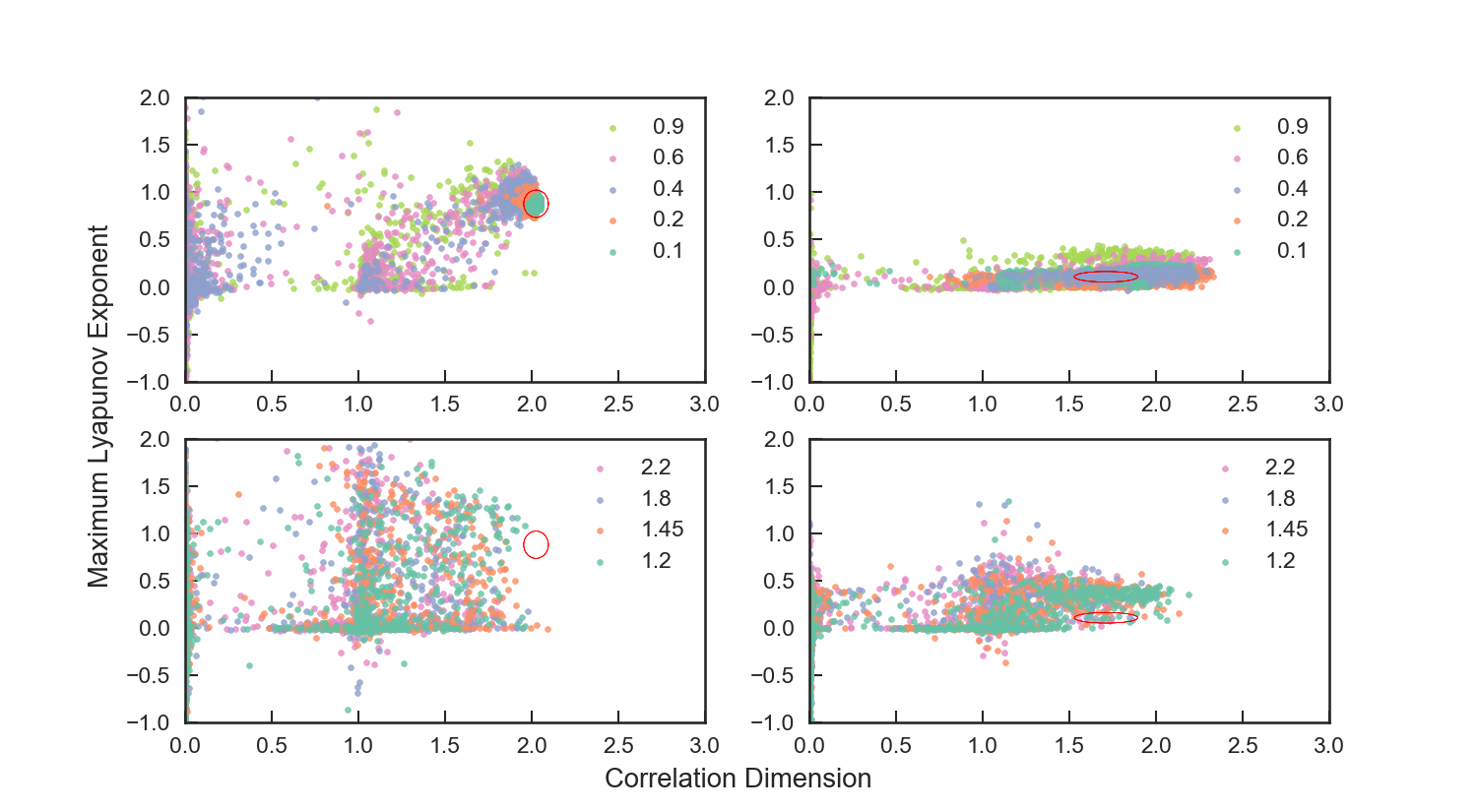}
    \caption{Largest Lyapunov exponent scattered against the correlation dimension for each of the $N=1000$ predictions per spectral radius. Results are shown for the Lorenz (left) and R{\"o}ssler system (right). Different spectral radii are differentiated by colours. The red object represents the five sigma error ellipse of both measures calculated based on 1000 simulated true trajectories.Random Erd{\"o}s-Renyi networks are used for the reservoir $\textbf{A}$.}
    \label{fig:corrlyap}
  \end{center}
\end{figure*}
%%%%%%
%http://www.dt.fee.unicamp.br/~tiago/courses/dinamica_caotica/Lyapunov.pdf shows for c=5.7 parametrisation leading to le=0.07 
\begin{table}[b!] 
\begin{center}
\begin{tabular}{ l | l | l}
    \hline
    Lorenz & Mean & $\sigma$ \\
    \hline
    \hline
    Correlation Dimension & 2.026 & 0.014 \\
    \hline
    Largest Lyapunov Exponent & 0.878 & 0.029 \\
    \hline
    \\
    \hline
    R{\"o}ssler & Mean & $\sigma$ \\
    \hline
    \hline
    Correlation Dimension & 1.713 & 0.037 \\
    \hline
    Largest Lyapunov Exponent  & 0.107 & 0.011 \\
    \hline
    \hline
\end{tabular}
\end{center}
\caption{\label{tab:meanstd} Mean and standard deviation $\sigma$ of the two measures calculated from 1000 simulated trajectories with different initial conditions.}
\end{table}
We simulate one trajectory which is used for the training of the network as well as for the comparison of the predicted trajectory with the actual one. Furthermore, we simulated additional $1000$ trajectories of the actual Lorenz and R{\"o}ssler system with different randomly chosen initial conditions in order to investigate the statistical error when calculating the correlation dimension and the largest Lyapunov exponent from the time series with limited length. 

Table~\ref{tab:meanstd} shows the means and standard deviations for both measures indicating that the error is reasonably small. As we use \textit{only} 10000 data points, our results are slightly below the expected values of around  $2.04$ for the correlation dimension and $0.89$ for the largest Lyapunov exponent of the Lorenz system. For the R{\"o}ssler system the results for the correlation dimension are significantly below the desired value of around $2$ because the Grassberger Procaccia algorithm is slower converging as compared to the Lorenz system when using only $10000$ data points. This is also reflected in the higher standard deviation $\sigma$ of the correlation dimension. However, we verified through increasing the number of data points that our calculations converge to the expected results. 

%%%%%%
Figure~\ref{fig:goodbad} shows two examples of predicted trajectories using reservoir computing in the setup described above with a spectral radius of $\rho=0.3$. Although we ran the prediction over $n=10000$ time steps we plotted the results for $n=2000$ time steps for the sake of clarity. On the left side of the plot one can see the trajectory of the $X$ coordinate for the predicted system using reservoir computing (blue) and the simulated system based on the Lorenz differential equations (green). In the upper plot both trajectories are overlapping for around 200 time steps and then deviate while still showing the characteristic pattern of the Lorenz system. However, in the lower plot both trajectories already separate after less than 100 time steps leading to a pattern which looks completely different. 

This is remarkable given the fact that the setup for both cases is identical except for a different random number seed which results in different realizations of the input function $\textbf{W}_{in}$ and the reservoir $\textbf{A}$. Since looking solely at the $X$ coordinate yields insufficient information about the overall prediction quality it is meaningful to investigate the whole attractor as plotted on the right side of Fig.~\ref{fig:goodbad}. Here we can see that the Lorenz attractor is reconstructed very well by the upper prediction while the lower prediction has nothing to do with the butterfly-shaped Lorenz attractor. Instead, the trajectory quickly runs into a fixed point after detaching and partly forming some kind of mirrored Lorenz attractor. The difference in prediction quality is not only reflected in the ability to match the original trajectory in the short-term but also with respect to the correlation dimension and the largest Lyapunov exponent. While in the upper case the resulting values of $\nu = 1.992$ and $\lambda = 0.851$ are well within the expectations for the Lorenz system, the lower realization completely misses to reconstruct the long-term climate. Immediately the question arises if this observation is an exception or whether the prediction quality is not robust with respect to different random number seeds. Therefore we systematically investigated this effect by running the same setup with $N=1000$ different random number seeds. Since it would be laborious to visually inspect the trajectories and attractors of all realizations we rely on the measures introduced in Section \ref{Methods} in order to assess if a prediction is good or bad. 

Figure~\ref{fig:corrfc} shows a scatter plot where the forecast horizon is plotted against the correlation dimension for all realizations. The colours represent different spectral radii and for each spectral radius there is one point for each of the $N=1000$ random number seeds. In order to make the results better readable we divided the plot into four sections where we grouped the results for four to five different spectral radii. The first thing we can observe is that the prediction of the Lorenz system using reservoir computing works better for smaller spectral radii. But more importantly one can also see that the prediction quality significantly varies even when using the same spectral radius - as already indicated by Fig.~\ref{fig:goodbad}. This becomes not only evident by the results for the forecast horizon but especially when considering the correlation dimension. Its values quickly spread when increasing the spectral radius indicating that the resulting predicted trajectories do not resemble the long-term climate of the system well in many cases. Figure~\ref{fig:corrfcRoessler} shows the same results for the R{\"o}ssler system. In contrast to the Lorenz system not the smallest spectral radius but a choice of $\rho = 0.4$ leads to the best results. In addition, the short-term prediction ability measured by the forecast horizon is significantly better with a number of predictions matching the original trajectory for 500 to almost 1000 time steps. However, the spread of the results for the correlation dimension seems to be larger as compared to the Lorenz system with high variability even for the best working spectral radii. This is partly due to the fact that the numerical calculation of the correlation dimension converges slower for the R{\"o}ssler system as mentioned in the previous section. As in Fig~\ref{fig:corrfc}, there seems to be some ``quantization" of the Forecast Horizon for both systems. The reason is that the predicted trajectory typically detaches from the actual one after completing a loop around the attractor.  

The variability becomes even clearer when looking at Fig.~\ref{fig:corrlyap}. Here we can see another scatter plot where the largest Lyapunov exponent is plotted against the correlation dimension and thus both components of assessing the long-term climate are reflected. The left plots shows the results for the Lorenz system where those for the R{\"o}ssler system are shown on the right side. In addition, the red ellipse shows the five sigma errors of the correlation dimension and the largest Lyapunov exponent calculated from 1000 simulations using the actual equations of the Lorenz and R{\"o}ssler system as shown in Table~\ref{tab:meanstd}. When studying the left side it becomes clear that even for the smallest and for the Lorenz system best working spectral radius $\rho=0.1$ (top left plot, dark green) the resulting "bubble" of points is of the same size as the $\sigma = 5$ error ellipse. This indicates strong variability given that $\sigma = 5$ is a large error. For the larger spectral radii shown in the bottom left plot - including the values $\rho=1.2$ and $\rho=1.45$ as used in Ref.~\cite{pathak2017using} - there is not a single point within the ellipse. This indicates that the prediction completely fails to reproduce the long-term climate for those cases. The results for the R{\"o}ssler system on the right side of the plot give a similar picture. However, even for the best working spectral radius of $\rho = 0.4$ there are many points outside of $\sigma = 5$ error ellipse. 
In addition, one can also see from the upper plots of Fig.~\ref{fig:corrlyap} that a good reproduction of the correlation dimension also tends to coincide with a better reproduction of the largest Lyapunov exponent although this effect is not very significant. 

\begin{figure}
  \begin{center}
    \includegraphics[width=1\linewidth]{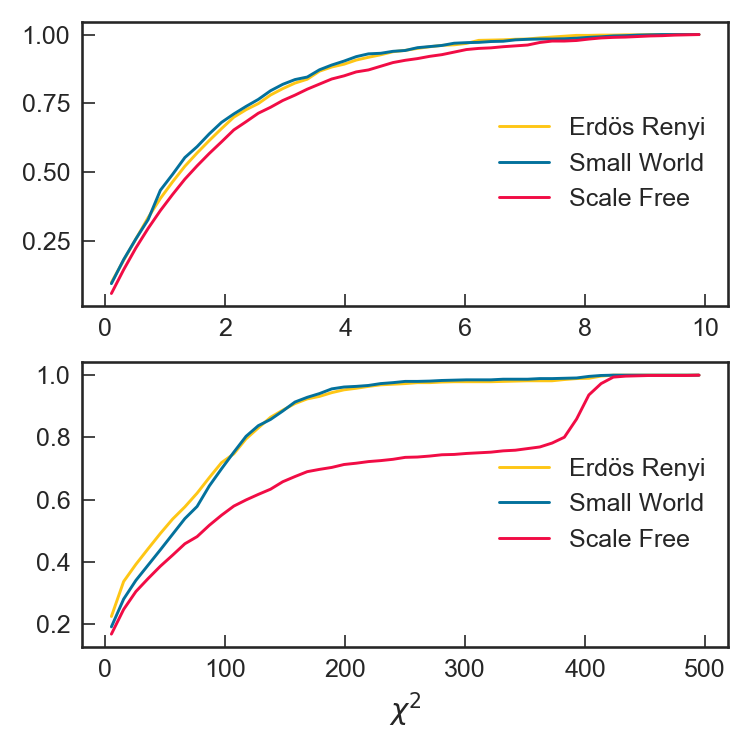}
    \caption{Top plot: Cumulative distribution of $\chi^{2}$ for the best working spectral radius $\rho=0.1$ of the Lorenz system calculated for values between 0 and 10. Bottom plot: Same for the R{\"o}ssler system with $\rho=0.4$ and values between 0 and 500}
    \label{fig:chisquared}
  \end{center}
\end{figure}
So far we only looked at the results based on the random Erd{\"o}s-Renyi networks. In order to compare the performance of the three different network topologies on a statistical level we perform a $\chi^{2}$ analysis of the long-term climate prediction. This means that we calculate
\begin{eqnarray}
\chi^{2}(i, \rho) =  \sum^{2}_{j=1} \left[\frac{X_{j}(i, \rho) - \langle X_{j} \rangle }{\sigma_{X_{j}}} \right] ^{2} \ , 
\label{eq:roessler}
\end{eqnarray}  
where $i$ is the $i-th$ random number seed, $\rho$ indexes the spectral radius. The sum goes over the correlation dimension ($j=1$) and the largest Lyapunov exponent ($j=2$). $\langle X_{j} \rangle$ represents the average value derived from $1000$ simulated actual Lorenz trajectories - as shown in Table~\ref{tab:meanstd} while $\sigma_{X_{j}}$ is the corresponding standard deviation. Therefore the resulting $\chi^{2}$ describes how strong the predicted results deviate from the actual values weighed by the errors of the actual values. 

Figure~\ref{fig:chisquared} shows the cumulative distribution of the $\chi^{2}$ for the three network topologies. The top plot contains the results for the Lorenz system using the spectral radius $\rho=0.1$ and evaluation values of $\chi^{2}$ for 0 and 10. We can see that Scale Free networks (red line) tend to work worst while Small World networks (blue) line slightly outperforms the Erd{\"o}s-Renyi networks (yellow line). However, it becomes clear that the method of reservoir computing works for all networks topologies tested here. In contrast, there is a difference between Scale Free networks and the other topologies in the case of the R{\"o}ssler system (bottom plot). Here, we calculated the cumulative distribution for values of $\chi^{2}$ between 0 and 500 based on $\rho=0.4$. The reason is that even for the best working spectral radius the variability is significantly higher as compared to the Lorenz system which leads to higher values of $\chi^{2}$ with only very few data points in the 0 to 10 range. It is interesting to note that the performance of Scale Free networks is now strongly below the other two networks while Erd{\"o}s-Renyi networks are slightly leading overall. Therefore, in contrast to the Lorenz system network topology seems to matter in this case. 

%%%%%%%%%%%%%%%%%%%%%%%%%%%%%%%%%%%%%%%%%%%%%%%%%%%%%%%%%%%%%%%%%%%%

\section{Conclusions and Outlook}
\label{Conclusion}
In this paper investigated the prediction of chaotic attractors by using reservoir computing from a statistical perspective. Instead of only predicting one trajectory we simulated 1000 realizations each -- where each realization corresponds to another random number seed -- for a number of different spectral radii $\rho$. Analyzing both the Lorenz and R{\"o}ssler system we found that the ability to exactly forecast the correct trajectory as well as the reconstruction of the long-term climate measured by correlation dimension and largest Lyapunov exponent strongly varies. Even for the exact same parameter setup there can be very good results that match the true trajectory for a large number of time steps and nicely reconstruct the attractor. On the other hand there can be results that completely fail in either one or both dimensions and are not reflecting the desired properties of the system. The results suggest that smaller spectral radii than typically used work better for both systems while in case of the Lorenz system even the smallest spectral radius $\rho = 0.1$ performed best. However, even in this case results show strong variability as they completely fill the five sigma error ellipse of correlation dimension and largest Lyapunov exponent. For the R{\"o}ssler system there are several predictions exceeding the $\sigma = 5$ error ellipse for the best working spectral radius of $\rho = 0.4$ and thus variability is even stronger as compared to the Lorenz system. This is an interesting observation given that the R{\"o}ssler system is considered less nonlinear. Overall our results indicate that special care must be taken in selecting the good predictions as predictions which deliver better short time prediction also tend to better resemble the long-term climate of the system.

Furthermore, we ran the same analysis using two other network topologies: Small world and scale free networks. In essence they produced comparable results with a slight outperformance of Small World networks and underperformance of Scale Free networks for the Lorenz system. For the R{\"o}ssler system the picture is different with a slight outperformance of Erd{\"o}s-Renyi networks while Scale Free networks are showing worse results in terms of the $\chi^{2}$ analysis. A tentative explanation for the lower performance of Scale Free networks could be the following: According to Singh \cite{singh2016scaling}, the more capable a brain or neuronal system is, the less scaling is present in its degree distribution. Overall it is important to point out that despite the differences presented here, the general methodology of reservoir computing works for different network topologies. However, even after trying different parameters and alternatively a setup where also the input states are going into the regression as described by Lukosevicius and Jaeger \cite{lukovsevivcius2009reservoir}, the variability can always be observed. 

% Next steps
Once the network is trained, the prediction is deterministic and depending strongly on the weights and only weakly on the topology. It should thus be possible to associate good and bad predictions with differential properties of the respective realization of the reservoir in a systematic way. First attempts in this direction for a reservoir with unweighted edges have recently been reported in Carroll and Pecora \cite{carroll2019network}. Once based on those insights more stable predictions are possible, a more precise analysis of the attractor properties e.g. with the spectrum of Lyapunov exponents could be useful and necessary. Furthermore, the role of the network size is also an interesting aspect. Current and future work is dedicated to the investigation of these questions - not the least because the answers to them will shed new light on the complexity of the underlying dynamical system.

%%%%%%%%%%%%%%%%%%%%%%%%%%%%%%%%%%%%%%%%%%%%%%%%%%%%%%%%%%%%%%%%%%%%%%%%%%%%%%%%
\section*{Acknowledgements}
We wish to acknowledge useful discussions and comments from Jonas Aumeier, Ingo Laut and Mierk Schwabe.

\bibliography{ReservoirComputing}

\end{document}